\documentclass{article}
\usepackage[utf8]{inputenc}
\usepackage{geometry}
\usepackage{amsmath,amssymb,graphicx,color,amsthm}
\usepackage[linesnumbered,ruled]{algorithm2e}
\usepackage{tikz}
\usepackage{dsfont}
\usepackage{mathtools}
\usetikzlibrary{shapes.geometric, arrows}
\usepackage{algorithmic}
\usepackage[english]{babel}
\usepackage[utf8]{inputenc}
\usepackage{mathrsfs}  
\usepackage{afterpage}
\usepackage{mathtools}
\usepackage{amsfonts}
\usepackage[colorinlistoftodos]{todonotes}
\usepackage{enumerate}
\usepackage[round]{natbib}
\usepackage{hyperref}
\usepackage{cleveref}
\usepackage{authblk}
\usepackage{graphicx}
\usepackage{amsmath}
\usepackage{multirow}

\usetikzlibrary{automata,positioning}
\usetikzlibrary{decorations.pathreplacing}

\newcommand{\excise}[1]{}

\newcommand\OO{\mathrm{O}}
\newcommand\RR{\mathbb{R}}
\newcommand\PP{\mathbb{P}}

\newtheorem{definition}{Definition}

\newtheorem*{example*}{Example}
\newtheorem{proposition}{Proposition}
\newtheorem{corollary}{Corollary}

\DeclareMathOperator\FRC{FRC}
\DeclareMathOperator\BFC{BFC}
\DeclareMathOperator\ORC{ORC}
\DeclareMathOperator\LRC{LRC}
\DeclareMathOperator\diam{diam}

\DeclareMathOperator\AER{AER}
\DeclareMathOperator\AOP{AOP}

\DeclarePairedDelimiterX{\infdivx}[2]{(}{)}{%
	#1\;\delimsize\|\;#2%
}

\newcommand{\RNum}[1]{\uppercase\expandafter{\romannumeral #1\relax}}

\makeatletter
\let\ams@underbrace=\underbrace
\def\underbrace#1_#2{%
  \setbox0=\hbox{$\displaystyle#1$}%
  \ams@underbrace{#1}_{\parbox[t]{\the\wd0}{\centering#2}}%
}
\makeatother

\xdefinecolor{dukeblue}{rgb}{0.004,0.129,0.412}
\xdefinecolor{uncblue}{rgb}{ 0.29411764705, 0.61176470588, 0.82745098039}

\title{Lower Ricci Curvature for Efficient Community Detection}
\date{}
\author[]{Yun Jin Park}
\author[]{Didong Li}
\affil[]{Department of Biostatistics, The University of North Carolina at Chapel Hill}
\graphicspath{{figures/}}

\begin{document}

\maketitle

\begin{abstract}
    This study introduces the Lower Ricci Curvature (LRC), a novel, scalable, and scale-free discrete curvature designed to enhance community detection in networks. Addressing the computational challenges posed by existing curvature-based methods, LRC offers a streamlined approach with linear computational complexity, making it well-suited for large-scale network analysis. We further develop an LRC-based preprocessing method that effectively augments popular community detection algorithms. Through comprehensive simulations and applications on real-world datasets, including the NCAA football league network, the DBLP collaboration network, the Amazon product co-purchasing network, and the YouTube social network, we demonstrate the efficacy of our method in significantly improving the performance of various community detection algorithms.
\end{abstract}

\section{Introduction}
In the modern era, the ubiquity of networks in various domains, from biological pathways~\citep{koutrouli2020guide} and social networks~\citep{ji2022co} to technological and cosmic webs~\citep{de2018network}, has fostered a significant interest in the study of complex systems. These networks, characterized by nodes representing entities and edges denoting interactions, provide a framework for understanding the intricate relationships and dynamics within these systems. Graph theory, applied to these network representations, has emerged as a vital tool for dissecting and interpreting the structural and functional intricacies of these interconnected systems~\citep{west2001introduction}.

Community detection is one of the most important aspects in the analysis of complex networks~\citep{dey2022community}. In these networks, communities represent subgroups of nodes (such as individuals, biological entities, or devices) that are more densely interconnected among themselves than with the rest of the network. The identification of these communities can yield invaluable insights into the structure and dynamics of the system being studied. For instance, in social networks, communities may represent groups of people with shared interests or connections, revealing patterns in social interactions and relationships~\citep{bakhthemmat2021communities}. In biological networks, such as those representing metabolic or protein-protein interaction pathways, community detection can help identify functional modules or clusters of interacting molecules, crucial for understanding biological processes and disease mechanisms~\citep{tripathi2019adapting}. Similarly, in technological networks, such as the internet or telecommunications networks, communities might consist of densely interconnected nodes or hubs that are critical for network functionality and resilience~\citep{zhang2022community}. By discerning these communities, we can gain a deeper understanding of not only the individual elements within the network, but also the overarching principles that govern their interactions and collective behavior.

Community detection methods have evolved significantly to address the diverse and complex structures of modern networks. Hierarchical Clustering~\citep{fortunato2010community, hastie2009elements}, for instance, has been instrumental in identifying nested community structures by iteratively merging or dividing groups based on their similarity. The Girvan-Newman algorithm~\citep{newman2004fast, newman2006modularity}, notable for its edge-betweenness centrality approach, has contributed substantially to understanding the modularity within networks. Similarly, Label Propagation algorithms~\citep{raghavan2007near}, recognized for their simplicity and speed, have been effective in detecting community structures in large networks by allowing nodes to adopt the majority label of their neighbors. The Walktrap algorithm~\citep{pons2005computing} has gained recognition for its approach of using random walks to identify communities, based on the idea that short random walks tend to stay within the same community. This method is particularly adept at capturing the local community structure in large networks. The Leiden algorithm~\citep{traag2019louvain}, an improvement over the well-known Louvain method~\citep{blondel2008fast}, offers enhanced accuracy and resolution in detecting communities. It addresses some of the limitations of previous methods by refining the community boundaries and ensuring a more balanced distribution of community sizes. These methods, each with their unique approaches and strengths, have collectively advanced our understanding of network structures, contributing to fields ranging from sociological studies to biological network analysis. 

A recent and significant development in network analysis is the discovery of a correlation between discrete curvature and community detection~\citep{sia2019ollivier}, which underscored the potential of using curvature-based methods to enhance our understanding and identification of communities within complex networks. Network curvature, particularly discrete curvature, has emerged as a powerful tool in the realm of graph theory and network analysis. The concept, rooted in geometric analysis, involves adapting the notion of Ricci curvature~\citep{ricci1900methodes,do1992riemannian}, traditionally applied to smooth manifolds, to discrete networks. This adaptation has led to the development of various discrete curvature measures, each offering unique insights into network properties.
 
One of the key forms of discrete curvature is the Ollivier-Ricci curvature (ORC, \cite{ollivier2007ricci}), which has been instrumental in studying transport efficiency and robustness in networks. It provides a measure of how the network deviates from a geometrically flat structure, offering insights into network connectivity and resilience. Another significant variant is the Forman Ricci curvature (FRC, \cite{forman2003bochner}), adapted from Riemannian geometry, which has been applied to analyze the shape and topological features of networks, proving useful in understanding the underlying structure of complex systems. The Balanced Forman curvature (BFC, \cite{topping2021understanding}), a refined version of the FRC, has been particularly effective in identifying bottleneck structures and critical connections within networks. This form of curvature is beneficial in applications where understanding the flow or distribution within a network is crucial. These curvature-based approaches have opened new avenues in network analysis, offering a geometric perspective to complement traditional topological and statistical methods. 
\begin{figure}[h!]
    \centering
    \includegraphics[width=0.8\textwidth]{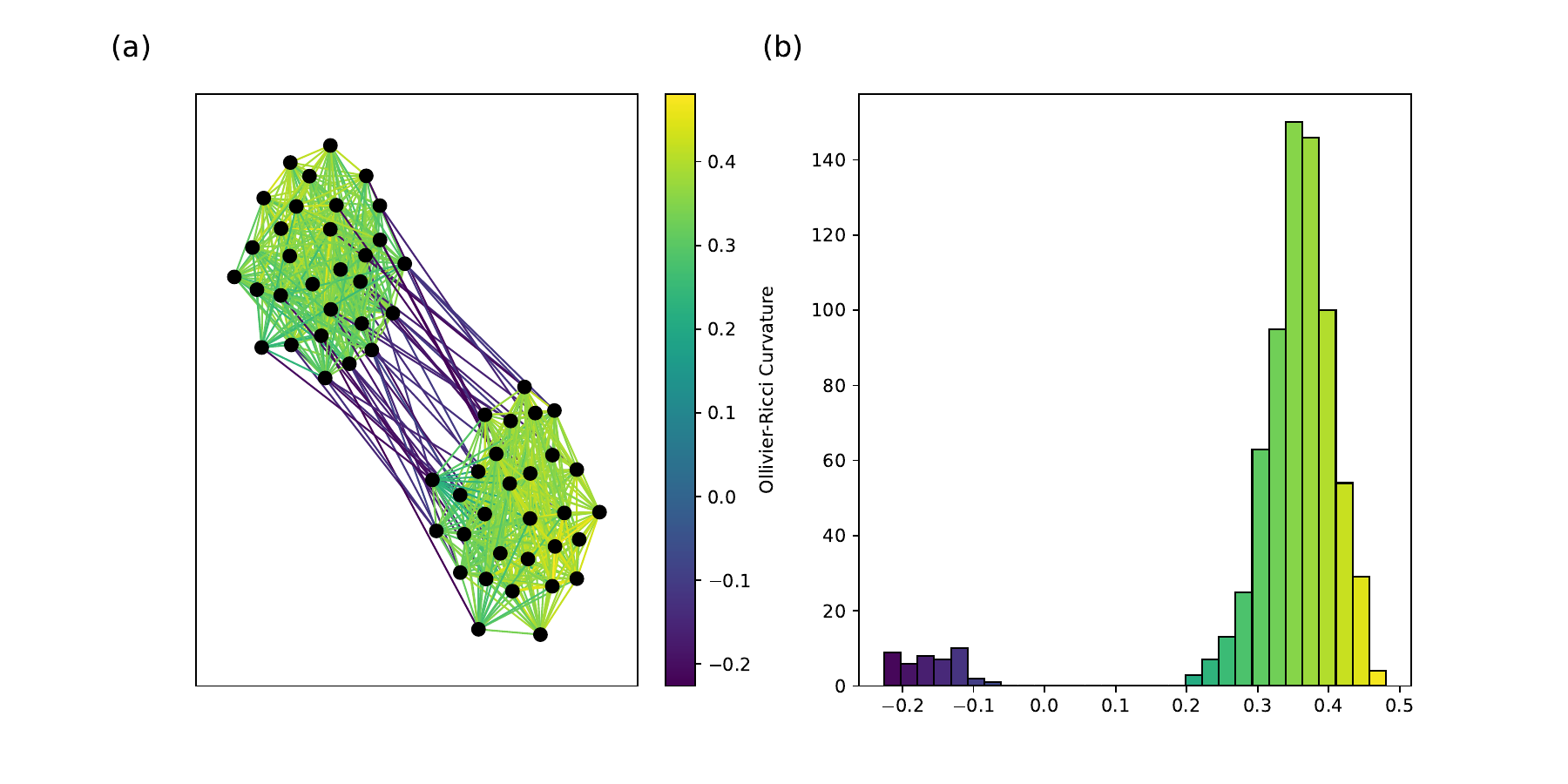}
    \caption{(a) A toy simulated network with two communities, with edges colored by ORC. (b) The histogram of ORC, suggesting its potential in community detection.}
    \label{fig:orchis}
\end{figure}

Although discrete curvature has found various applications in network analysis, such as understanding internet topology~\citep{ni2015ricci}, differentiating cancer networks~\citep{sandhu2015graph}, and addressing oversquashing and oversmoothing problems in graph neural networks~\citep{nguyen2023revisiting}, its specific use in community detection remains relatively underexplored. A notable exception is \cite{sia2019ollivier}, who proposed using ORC for community detection. Figure \ref{fig:orchis} shows a toy network where two distinct communities are apparent. In this network, the edges connecting different communities tend to have lower ORC values, while those within a community exhibit higher ORC values. The proposed method involves iteratively removing the edge with the smallest ORC and recalculating all edge ORCs until the network becomes disconnected, with each connected component identified as a community. Similarly, \cite{fesser2023augmentations} proposed another iterative algorithm to remove edges with augmented FRC above a threshold, until no edge curvature exceeds that threshold. However, these approaches have several major drawbacks. First, the computational cost of calculating ORC and augmented FRC is high, scaled as $\OO(mn^3)$ for ORC and $\OO(mn^2)$ for augmented FRC, where $m$ represents the number of edges and $n$ the number of nodes. This makes it prohibitively expensive for large scale such as the DBLP co-authorship network ($n=317,080,~m=1,049,866$), the Amazon product co-purchasing network~($n=334,863,~m=925,872$), and the YouTube social network ($n=1,134,890, ~m=2,987,624$, \cite{yang2012defining}). Second, the iterative nature of the algorithm introduces significant extra time inefficiencies. In the worst-case scenario, up to $\OO(m)$ iterations might be required. Third, the methods, despite their innovative approach, can sometimes be too restrictive and may underperform compared to popular methods such as the Leiden algorithm. 

To effectively tackle the challenges in community detection within large-scale networks, our study introduces a novel curvature measure, the Lower Ricci Curvature (LRC). LRC is specifically designed for efficient computation, with a linear computational complexity of $\OO(mn)$. This significantly reduces the computational burden compared to traditional curvature measures, making it highly suitable for large networks.
In addition to its computational efficiency, we provide some theoretical analysis of LRC, particularly its connection to the Cheeger constant, a well-established concept in graph theory~\citep{mohar1989isoperimetric}, which helps in understanding how LRC relates to the division of a network into communities. 

Building on the theoretical foundation of LRC, we have developed a preprocessing algorithm that utilizes LRC to improve existing community detection methods. This algorithm is designed to be suitable for a wide range of applications, due to its adaptability to different network structures and sizes and its compatibility with various community detection methods. Our approach was rigorously tested through both simulation studies and real-world data analysis. We applied it to networks of diverse sizes, including both small-scale networks (NCAA football league network) and large-scale networks with mixed membership (the DBLP coauthorship network, the Amazon product co-purchasing network, and the YouTube social network). The results from these studies consistently demonstrate that our preprocessing step, based on the Lower Ricci Curvature, not only enhances the efficiency but also improves the accuracy of various established community detection methods.

Proofs and additional experimental details are provided in the Appendix.

\section{Background}
\subsection{Community detection}

Community detection in network analysis is essential for unraveling the intricate structures of networks. Communities are typically defined as subgroups of nodes with denser internal connections compared to their external connections \citep{radicchi2004defining}. Understanding these communities is vital to reveal the main structural characteristics of networks and to classify nodes based on their interrelations \citep{fortunato2016community}.

While the Introduction briefly mentions various community detection algorithms, this subsection aims to delve deeper into their specific functionalities and contributions. The Girvan-Newman algorithm leverages edge betweenness centrality and hierarchical clustering to identify community structures \citep{newman2004finding}. The Leiden algorithm, evolving from the Louvain method, focuses on optimizing modularity, thereby enhancing the resolution and accuracy of detected communities \citep{blondel2008fast, traag2019louvain}. Other notable methods include Label Propagation, which relies on the diffusion of information \citep{raghavan2007near}, and Walktrap, which uses random walks to discern community structures \citep{pons2005computing}.

Additionally, algorithms such as the Angel \citep{rossetti2020angel}, ego-based community detection (Ego, \cite{leskovec2012learning}), K-clique~\citep{palla2005uncovering}, Speaker-Listener Label Propagation Algorithm (SLPA, \cite{xie2011slpa}) contribute diverse perspectives and techniques to community detection. Each of these methods brings unique strengths to the analysis of network structures, addressing different aspects and challenges in identifying community patterns.

To evaluate the performance of these algorithms, criteria such as the Adjusted Rand Index (ARI) and the Adjusted Mutual Information (AMI) are commonly used. ARI evaluates the agreement in node pair assignments between clustering results, offering a quantitative assessment of similarity \citep{rand1971objective}. AMI measures the similarity between different community detection results on the same dataset, providing insights into the amount of shared information \citep{vinh2009information}.  In our study, we utilize these criteria to gauge the improvements in community detection algorithms' performance when incorporating our newly proposed LRC-based preprocessing step.

\subsection{Discrete curvatures} 

Curvature, a fundamental concept in mathematics, describes how a curve deviates from a straight line or a surface from being flat~\citep{boothby1986introduction}. In Riemannian geometry, curvatures such as Ricci curvature, provide insights into the unique geometry properties of different spaces, including volume changes along geodesics~\citep{do1992riemannian}. Historically, generalizing this concept to discrete objects, such as networks, presented a significant challenge. A pivotal moment in this endeavor came with the work of Forman~\citep{forman2003bochner}, who innovatively adapted curvature concepts to the discrete realm. This milestone opened the door for further exploration and application of curvatures in discrete spaces, including networks. 

For presentational simplicity, in this paper, we focus on an unweighted graph $G = (V, E)$, where $V$ is a set of nodes and $E$ is a set of edges, but the framework can be generalized to a weighted graph in a straightforward manner. Let $(ij)$ be an edge connecting node $i$ and node $j$, we denote the degree of $i$, i.e., the number of edges of node $i$, by $n_i$, the number of shared neighbors of $i,j$, i.e., the number of triangles based on $(ij)$, by $n_{ij}$. Under this framework, the BFC is defined as follows:

\begin{definition}[Forman Ricci Curvature (FRC)]
The FRC of edge $(ij)$ is defined as 
$$\FRC(ij) = 4 - n_i - n_j + 3n_{ij}.$$
\end{definition}

The computational cost for calculating FRCs for all edges is $\OO(mn)$. Following Forman's groundbreaking work, there has been a surge of studies exploring and applying what is now known as the Forman Ricci curvature, or FRC, to various network structures. For example, \cite{sreejith2016forman_directed} extends FRC from undirected to directed networks, and \cite{sreejith2016forman} extends FRC to complex networks.

However, its unbounded and scale-dependent nature, as well as its skewness toward negative values in various networks pose interpretational challenges~\citep{sreejith2016forman}. To address these issues, an improved version known as the balanced Forman curvature (BFC, \cite{topping2021understanding}) was proposed:
\begin{definition}[Balanced Forman Curvature (BFC)]
The BFC of edge $(ij)$ is defined as 
$$\BFC(ij) = \frac{2}{n_i} + \frac{2}{n_j} -2 + 2\frac{n_{ij}}{\max(n_i, n_j)} + \frac{n_{ij}}{\min(n_i, n_j)} + \frac{s_{i,j} + s_{j,i}}{\gamma_{\max}\max(n_i, n_j)},$$
where $s_{i,j}$ is the number of neighbors of node $i$ forming a 4-cycle based at the edge $(ij)$ without diagonals inside, $\gamma_{\max}$ is the maximal number of 4-cycles based at edge $(ij)$ traversing a common node (see \citep{topping2021understanding} for more details). 
\end{definition}
 BFC, with its bounded range $[-2,2]$, has been applied to identify the bottleneck structure in graphs, particularly for addressing the over-squashing phenomenon in graph neural networks. However, its computational complexity of $\OO(mn^2)$, limits its application to large-scale networks, mainly due to computationally intensive terms $s_i$ and $\gamma_{\max}$, which involves counting the number of squares based at nodes $i$ and $j$ under certain constraints~\citep{topping2021understanding}. 

Simultaneously, Ollivier made significant contributions by defining the Ricci curvature for networks through optimal transport and differential equations~\citep{ollivier2007ricci}:

\begin{definition}[Ollivier-Ricci Curvature (ORC)]
The ORC of edge $(ij)$ is defined as
$$\ORC(ij) = 1 - \frac{W_1(m_i, m_j)}{d(i,j)},$$
where $W_1$ is Wasserstein-1 distance, $m_i$ is a local probability measures at node i, defined as 
\begin{equation}
    m_i(k)=
    \begin{cases}
      \frac{1}{n_i}, &(ik)\in E\\
      0, & (ik)\notin E
    \end{cases}
  \end{equation}
 and $d(i, j)$ is the length of a shortest path from node $i$ to node $j$, also known as the graph distance. 
\end{definition}

Ollivier-Ricci curvature, or ORC, has sparked a wide array of follow-up research, further enriching the field of network analysis with these novel curvature-based insights. For example, \cite{lin2010ricci,lin2011ricci,erbar2012ricci,bauer2013generalized} have provided deep mathematical insights into the properties and implications of ORC in the context of graph theory and network geometry. Notably, \cite{sia2019ollivier} utilized ORC for community detection, iteratively removing the edge with the smallest ORC. However, its computational cost ($\OO(mn^3)$) poses significant challenges, especially for iterative algorithms. 

\subsection{Stochastic Block Model (SBM)}
To illustrate network curvatures in a simplified context, we consider the Stochastic Block Model (SBM) in this manuscript, a basic yet versatile model used in network analysis~\citep{holland1983stochastic}. SBM is renowned for its ability to mimic community structures within networks. In this model, nodes are partitioned into $K$ distinct communities, and connections between nodes are probabilistically determined based on their community memberships.

Each node $i$ is assigned a community label $z_i\in\{1,\cdots,K\}$, indicating its community membership. The block matrix $B$, a $K\times K$ symmetric matrix, is a critical component of the SBM, dictating the probability of edge formation between nodes from communities. Specifically, $B_{kl}$ represents the probability of an edge existing between nodes from community $k$ and community $l$. In an SBM, the probability of an edge existing between any two nodes $i$ and $j$ follows a Bernoulli distribution, and is independent of other edges, as reflected in the adjacency matrix $A\in\RR^{n\times}$ :
$$\PP(A_{ij}=1)=B_{z_iz_j}$$

While a basic SBM might appear too simplistic for complex real-world data, its extensions, such as hierarchical SBM and mixed membership SBM, offer more nuanced representations. These models cater to scenarios involving hierarchical community structures~\citep{peixoto2014hierarchical} or nodes with memberships in multiple groups~\citep{airoldi2008mixed}.

\section{Lower Ricci Curvature (LRC)}
In this section, we introduce a novel discrete curvature, the Lower Ricci Curvature (LRC), notably for its high performance in community detection and low computational complexity of $\OO(mn)$. We delve into the intuition behind defining LRC, and how these factors contribute to its computational efficiency and efficacy in community detection.
 
The LRC of edge $ij$ is defined as:
$$\LRC(ij) = \frac{2}{n_i} + \frac{2}{n_j} -2 + 2\frac{n_{ij}}{\max(n_i, n_j)} + \frac{n_{ij}}{\min(n_i, n_j)}.$$
Several key observations about LRC can be made. Firstly, the computation of $\LRC(ij)$ requires $\OO(n)$, similar to FRC. Secondly, LRC is always within the range of $[-2,2]$, aligning with the bounds of BFC. Third, LRC is consistently less than or equal to BFC, with the difference, denoted as $\Delta(ij)$, defined as
$$\Delta(ij)\coloneqq \frac{s_i + s_j}{\gamma_{\max}\max(n_i, n_j)}=\BFC(ij)-\LRC(ij)\geq 0.$$
In fact, BFC is further upper bounded by ORC~\citep{topping2021understanding}, leading to the following proposition, which underlines why we term it \textit{lower} Ricci curvature.
\begin{proposition}\label{prop:bounds}
For any edge $(ij)$, 
$$\LRC(ij)\leq \BFC(ij)\leq \ORC(ij).$$
\end{proposition}

This proposition motivates our first rationale for defining LRC.  The computational bottleneck of BFC is the term $\Delta$, which requires $\OO(n^2)$ time. However, this leads to two pertinent questions. First, is LRC effective in differentiating edges within and between communities? Second, does the term $\Delta$ significantly contribute to community detection, or is there a notable difference in $\Delta$ for edges within the same community versus those between different communities? 

\begin{figure}[h!]
    \centering
    \includegraphics[width=\textwidth]{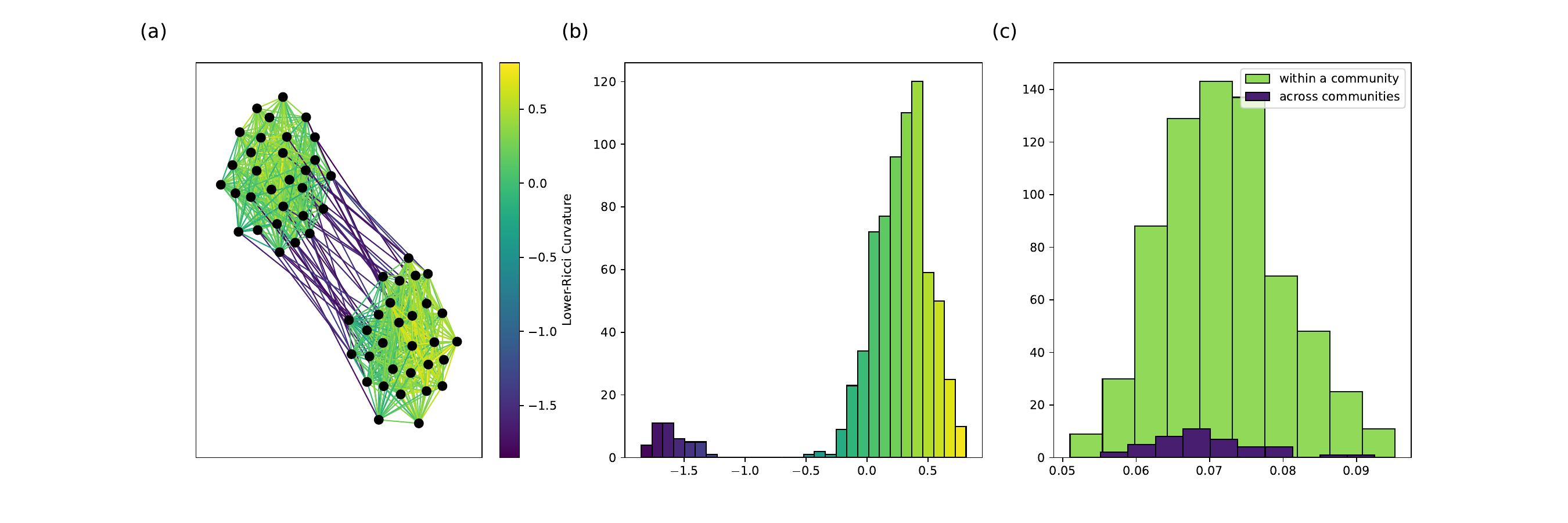}
    \caption{(a) A SBM-generated network with $K=2$, $B_{kk}=0.8$, $B_{kl}=0.05$ for $k\neq l$, with edges colored by LRC. (b) The histogram of LRC, suggesting its potential in community detection. (c) The histogram of $\Delta$ for within and across community edges, indicating that $\Delta$ may not significantly contribute to community detection.}
    \label{fig:lrcex}
\end{figure}

To further investigate these questions, we utilize an SBM-generated network as a toy example. Figure \ref{fig:lrcex} presents a network with $n=60$ nodes, divided evenly into two communities. Edges within communities have a higher probability of $0.8$, while edges between communities are set at a lower probability of $0.05$. The edges are color-coded based on their LRCs: higher LRCs are marked in yellow, while lower LRCs are marked in purple. This visual representation helps highlight that edges bridging different communities tend to have smaller LRC values compared to those within the same community.
This example not only serves as a visualization exercise but also supports the use of LRC in community detection. It demonstrates that LRC achieves computational efficiency by omitting the computationally expensive term $\Delta$ without sacrificing its ability to detect community structures.

The direct link between LRC and ORC is less straightforward, which guides the second intuition behind the definition of LRC. The primary computational challenge in calculating ORC lies in the Wasserstein-1 distance, also known as the earth moving distance~\citep{villani2009optimal}. A natural approach is to approximate this distance or bound it from below, above, or both. To effectively bound ORC, it is necessary to bound the Wasserstein distance, which involves complex calculations of the total cost of certain candidate transports (see \cite{jost2014ollivier} for more details). The bounds are established as follows:

$$\frac{2}{n_i} + \frac{2}{n_j} -2 + 2\frac{n_{ij}}{\max(n_i, n_j)} + \frac{n_{ij}}{\min(n_i, n_j)}\leq \ORC(ij)\leq \frac{n_{ij}}{\max\{n_i,n_j\}}.$$
Notably, the lower bound is precisely the LRC, which connects back to Proposition \ref{prop:bounds}. While it is technically feasible to use the upper bound, the focus is on the lower bound, i.e., LRC, due to its practical performance and the positivity of the upper bound.

As a direct corollary of these inequalities, the following corollary establishes a link between the bound of LRC and the diameter of the network, defined as $\mathrm{diam}(G)=\sup_{i,j\in V} d(i,j)$, where $d$ represents the graph distance. This is also related to the Cheeger constant \citep{chung1997spectral}, a measure indicative of the presence of a community structure in the network.

\begin{corollary}\label{cly:Cheeger}
If there exists $\alpha>0$ such that $\LRC(ij)\geq \alpha$ for any $(ij)\in E$, then
\begin{enumerate}
    \item $\diam(G)\leq \frac{2}{\alpha}$.
    \item $\frac{\lambda_1}{2}\geq h_G\geq \frac{\alpha}{2}$, where $\lambda_1$ is first non-zero eigenvalue of the normalized graph Laplacian, also known as the spectral gap, and $h_G$ is the Cheeger constant.  
\end{enumerate}
\end{corollary}
The interpretation of this corollary is that a larger value of $\alpha$, suggests a graph structure more akin to a fully connected graph, hence a smaller diameter. Similarly, a larger $\alpha$ correlates with a higher Cheeger constant, indicating a more interconnected network with less pronounced separability into distinct community structures.


We conclude this section with a comparative overview of the four curvatures: FRC, BFC, ORC, and LRC. The key to this comparison is their computational complexity and whether they are scale-free, i.e., independent of the network size characterized by $n$ and $m$. Scale-free properties are particularly important in the network analysis, as they ensure the applicability and consistency of curvature measures across networks of different sizes. This quality is preferable as it allows for meaningful comparisons and generalizations across various network structures, from small-scale to large-scale networks, without being biased by their size. 

\begin{table}[!h]
\centering
\begin{tabular}{|c|c|c|}
\hline
Curvature & Computational Complexity & Scale-Free\\
\hline
FRC & { $\bf O(mn)$} & No\\
\hline
BFC& $\OO(mn^2)$ & {\bf Yes}\\
\hline
ORC & $\OO(mn^3)$ & {\bf Yes}\\
\hline
LRC & { $\bf O(mn)$} & {\bf Yes}\\
\hline
\end{tabular}
\caption{Comparison of four curvatures.}
\label{tab:curvatures}
\end{table}
Among these curvatures, LRC stands out for its linear computational complexity and scale-free property, making it a versatile and efficient choice for network analysis. This blend of computational efficiency and scale-free nature makes LRC an ideal candidate for analyzing networks in various contexts.

\section{LRC-based preprocessing }
As observed in \Cref{fig:lrcex}(b), the presence of community structures in networks often results in a bimodal distribution of LRC values. This typically manifests itself as two distinct modes in the histogram of LRCs: a smaller mode corresponding to across-community edges and a larger mode representing within-community edges. This observation underpins our proposed preprocessing step for community detection: removing edges with small LRC values below a specific threshold. This approach aims to retain more within-community edges, thereby making the community structure more pronounced. The threshold is determined by fitting a Gaussian mixture model (GMM, \cite{reynolds2009gaussian}) to the histogram of LRCs, as outlined in the following algorithm.

\begin{figure}[h!]
    \centering
    \includegraphics[width=\textwidth]{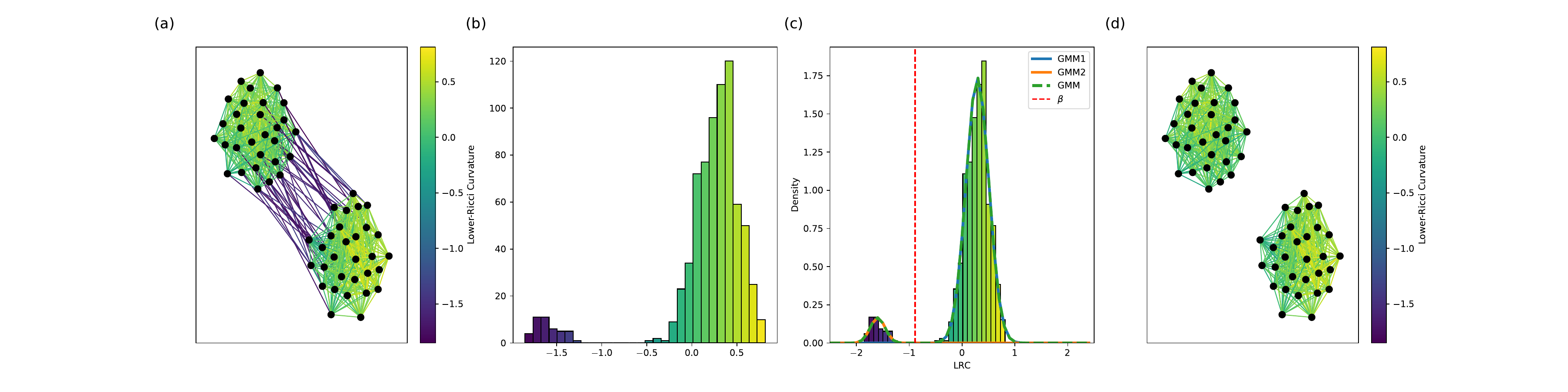}
    \caption{(a) A SBM-generated network with $K=2$, $B_{kk}=0.8$, $B_{kl}=0.05$ for $k\neq l$, with edges colored by LRC. (b) The histogram of LRC, with each bar colored by LRC. (c) The threshold $\beta$ (the dotted vertical line) estimated by GMM. GMM1 is the mixing component with a large mean $\mu_2$, GMM2 is the mixing component with a smaller mean $\mu_1$. (d) The processed network, exhibiting a more discernible community structure.}
    \label{fig:pre}
\end{figure}

\begin{algorithm}[H]
\caption{LRC-based preprocessing algorithm for community detection}\label{alg:cap}
  \KwIn{Raw network data: $G = (V, E)$}
  \KwOut{Preprocessed network data $G' = (V, E')$}
  Calculate the LRC for all edges\;
  Fit a Gaussian mixture model with two mixing component to LRCs, obtaining the estimate $\hat{p}(x)=\pi_1 N(x;\mu_1,\sigma_1^2)+\pi_2 N(x;\mu_2,\sigma_2^2)$, where $\mu_1<\mu_2$\;
  Find the local minimum $\displaystyle{\beta\coloneqq \inf_{\mu_1<x<\mu_2}\hat{p}(x)}$\;
  Remove all edges with LRCs smaller than $\beta$: $E'\coloneqq \{(ij)\in E:\LRC(ij)\geq \beta\}$
\end{algorithm}
The workflow of our proposed method is illustrated in \Cref{fig:pre}, based on the example network previously discussed.
This toy example illustrates how our preprocessing step is expected to enhance the performance of existing community detection methods by clarifying the underlying community structures.

Following the description of the underlying community structures, it's crucial to highlight the efficiency and scalability of the LRC-based preprocessing method. The calculation of LRC itself requires $\OO(mn)$ time, and the subsequent edge removal step is a one-time, non-iterative process. This is in stark contrast to competitor methods that rely on iterative processes~\citep{jost2014ollivier,sia2019ollivier}, which can significantly increase computational time, especially for large networks that are increasingly common in various domains. 

Crucially, this increase in efficiency does not compromise the accuracy of community detection. In the following sections, we present empirical evidence showing how our LRC-based preprocessing not only maintains but often enhances the effectiveness of popular community detection algorithms, even in complex network scenarios. This demonstrates the dual benefit of our approach: it streamlines computation while enriching the depth of network analysis.

\section{Simulation}

To access the effectiveness of LRC in community detection, we conducted simulations using networks generated from SBM. For a pair of  within community edge probability $p_1$ and across community edge probability $p_2$ (with $p_2<p_1$), we generate 100 graph replicates, each with $n=100$ nodes. We evaluated three distinct scores, motivated by our proposed preprocessing method, to compare the performance of LRC against three other existing curvatures. The results are visually represented through heat maps, with the x-axis representing $p_1$, the y-axis representing $p_2$, and the color indicating the score. 

\begin{enumerate}
\item \textbf{Proportion of Perfect Separation (PPS).} The first score we considered is the Proportion of Perfect Separation. For each graph replicate, we calculated the minimum curvature value among within-community edges and the maximum curvature value among across-community edges. A situation where the minimum within-community curvature exceeds the maximum across-community curvature indicates perfect separation of within and across community edges. This implies that our preprocessing would remove all across-community edges while retaining all within-community edges, enabling effective community detection by any reasonable method postprocessing. Mathematically, PPS is the proportion of networks satisfying $\inf_{(ij)\in E,z_i=z_j}\LRC(ij)\geq \sup_{(ij)\in E,z_i\neq z_j}\LRC(ij)$. The score ranges between 0 and 1, with higher values indicating better performance.

\begin{figure}[h!]
  \centering
    \includegraphics[width=\textwidth]{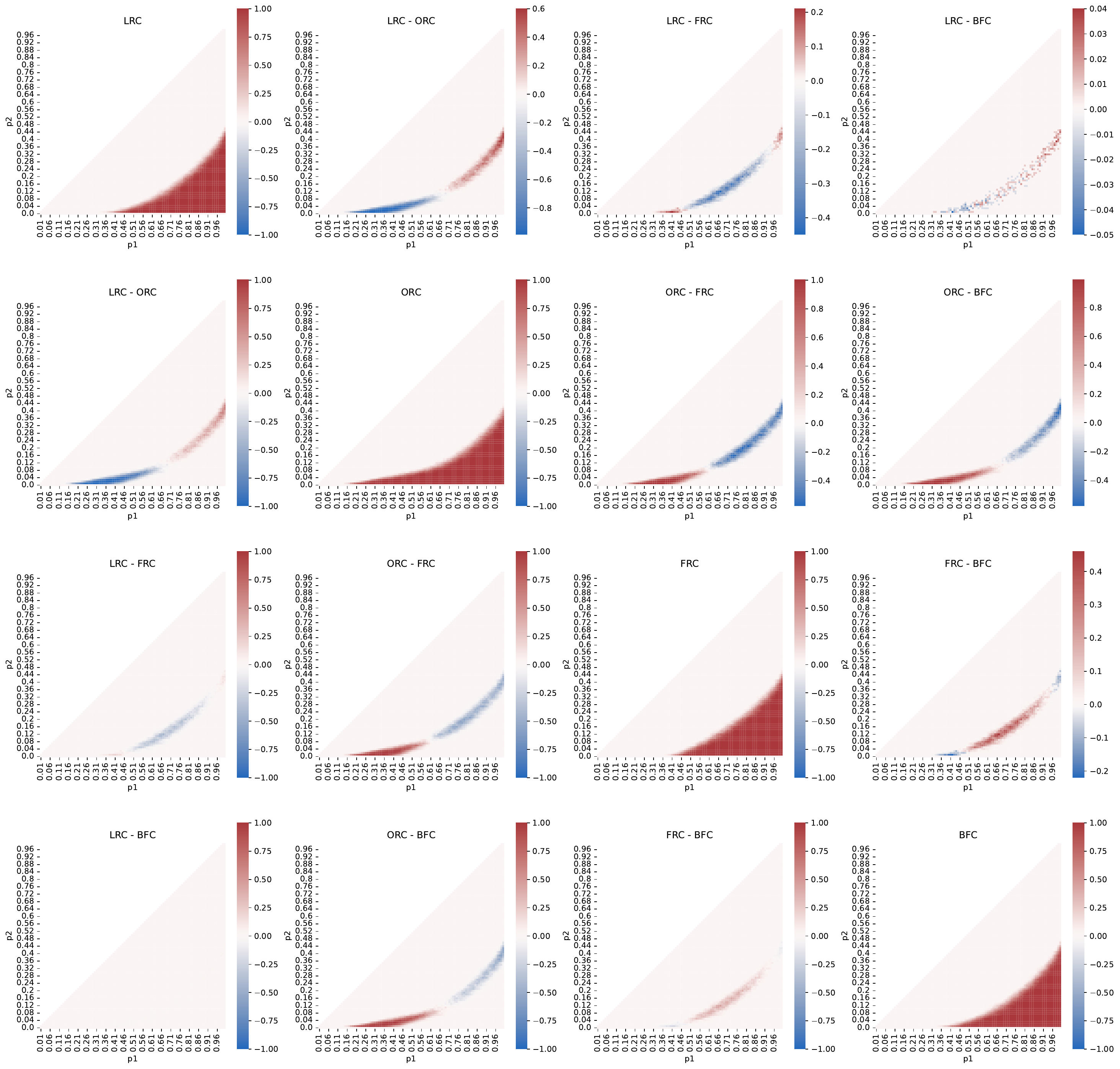}
    \caption{Comparison heat map for PPS. }
    \label{fig:PPS}
\end{figure}

The diagonal heat maps in Figure \ref{fig:PPS} depict the extent of separation between within-community and across-community curvature distributions. A redder hue indicates a higher degree of separation. These maps suggest that all the four curvatures, including LRC, effectively differentiate community structures across a variety of \( p_1, p_2 \) pairs. The off-diagonal heat maps in the lower triangle compare the performance of each curvature with others (red for superior performance, blue for inferior), with a raw scale of \([-1, 1]\). The upper triangle heat maps also compare curvature performances, but with normalized ranges to amplify differences. Overall, LRC shows comparable performance in PPS compared to other curvatures.

\item {\bf Average within-community Edge Removal Ratio (AER)} The second score, AER, provides a softer evaluation compared to PPS. While PPS focuses on perfect separation, AER quantifies the extent to which within-community edges might be incorrectly removed when aiming to eliminate all across-community edges. This score is particularly insightful, as it accounts for the potential drawback of our preprocessing method in mistakenly removing valuable within-community connections.

Mathematically, AER is defined as the ratio of the number of within-community edges, whose LRC values are smaller than the maximum LRC value of across-community edges, to the total number of within-community edges. In formula terms, AER is given by the proportion 
$$\AER \coloneqq \frac{\left|\left\{(ij)\in E,z_i=z_j:\LRC(ij)<\sup_{(ij)]\in E,z_i\neq z_j}\LRC(ij)\right\}\right|}{\left|\left\{(ij)\in E,z_i=z_j\right\}\right|}.$$
A score of $0$ indicates ideal performance (no within-community edges are incorrectly removed), aligning with a PPS of $1$. Conversely, an AER of $1$ implies the extreme scenario where all edges are erroneously removed.

Figure \ref{fig:AER} presents the heat map of AER scores, organized similarly to the PPS heat map. The diagonal panels show the AER score for each curvature, while the off-diagonal panels compare the performance of different curvatures using the AER score. A visualization in these heat maps can provide insights into how effectively each curvature avoids the unintended removal of within-community edges, which is crucial for maintaining the integrity of the community structure during preprocessing.
\begin{figure}[h!]
  \centering
    \includegraphics[width=\textwidth]{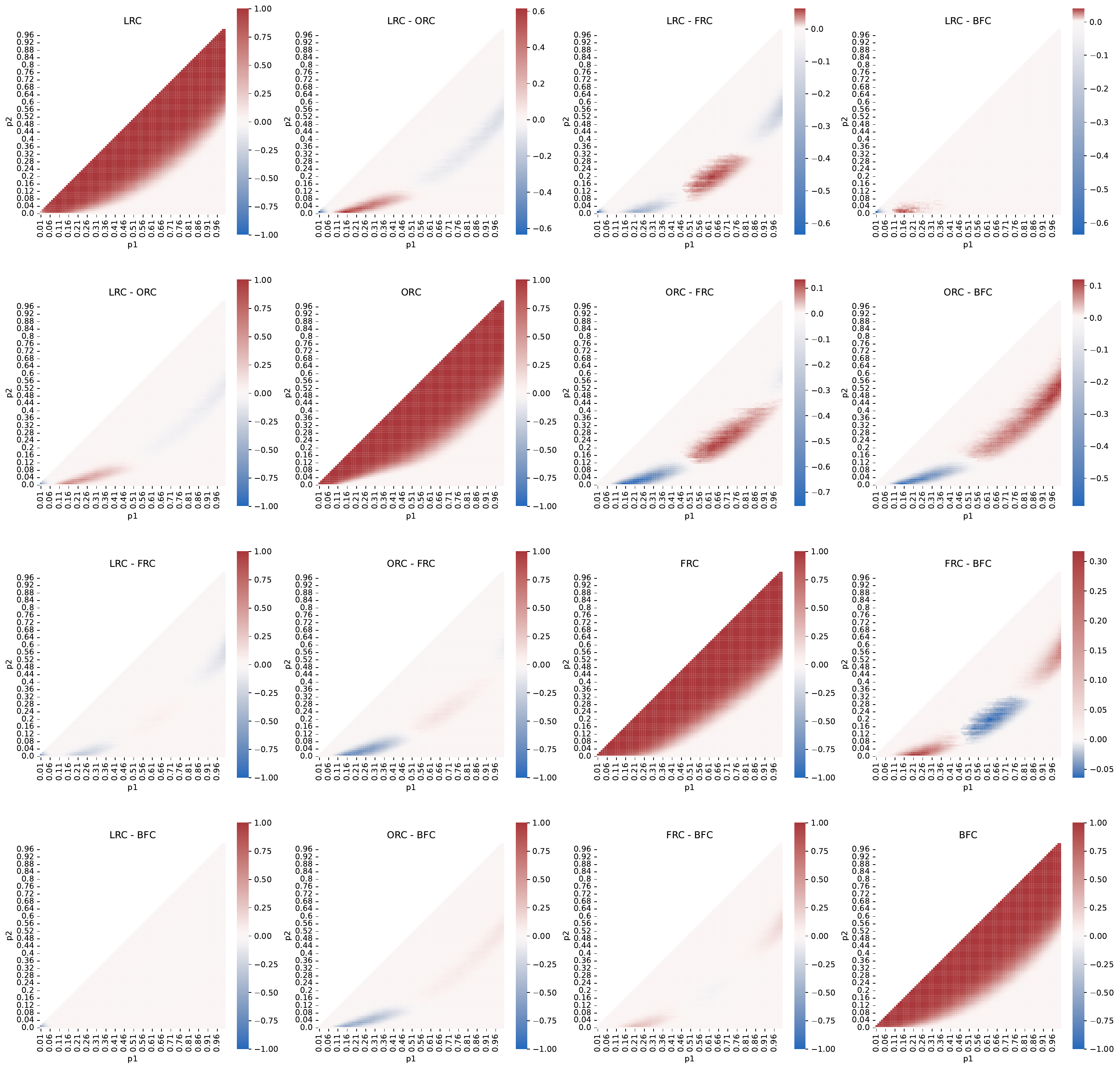}
    \caption{Comparison heat map for AER}\label{fig:AER}
\end{figure}

\item {\bf Average overlapping percentiles {AOP}.} The third score, AOP, is designed to quantify the extent of overlap between the curvature distributions of within-community and across-community edges in a more symmetric manner. This score captures the degree to which these two distributions intermingle, providing a nuanced view of the effectiveness of a curvature in distinguishing community structures.

Mathematically, AOP is calculated as follows: For each replicate graph, we determine the percentile of the minimum within-community LRC value within the distribution of across-community LRC values. We then calculate one minus the percentile of the maximum across-community LRC value within the distribution of within-community LRC values. The AOP score is the sum of these two quantities. Formally, it can be expressed as:
$$\AOP\coloneqq \mathscr{P}_{\inf_{(ij)\in E,z_i=z_j}}\left(\left\{\LRC(ij):(ij)\in E,z_i\neq z_j\right\}\right)+1-\mathscr{P}_{\sup_{(ij)\in E,z_i\neq z_j}}\left(\left\{\LRC(ij):(ij)\in E,z_i= z_j\right\}\right),$$
where $\mathscr{P}_aA$ is the $a$-th percentile of set A. 

In the ideal scenario where there is no overlap, the first percentile would be 1 (indicating the minimum within-community LRC is at the highest end of the across-community distribution), and the second percentile would also be 0 (indicating the maximum across-community LRC is at the lowest end of the within-community distribution), resulting in an AOP score of 2. Conversely, in the worst-case scenario where there is complete overlap, the AOP score becomes 0.
\begin{figure}[h!]
  \centering
    \includegraphics[width=\textwidth]{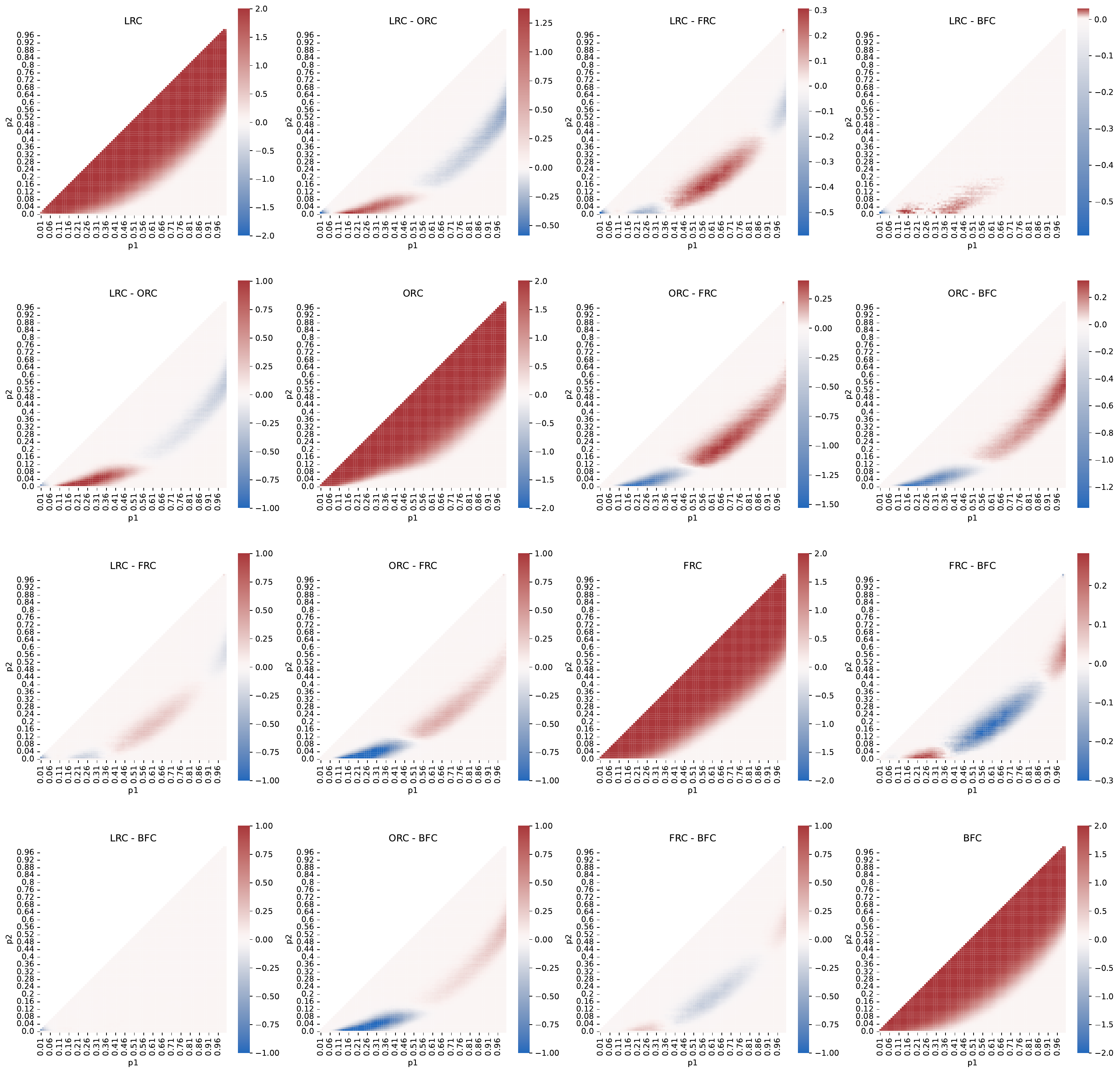}
    \caption{Comparison heat map for AOP.}
    \label{fig:AOP}
\end{figure}
Figure \ref{fig:AOP} displays the heat map visualization of AOP scores, arranged similarly to the previous scores. The diagonal panels show the AOP for each curvature, while the off-diagonal panels compare different curvatures using the AOP measure. This visualization aids in understanding the extent to which each curvature can differentiate community structures by evaluating the overlapping of curvature distributions.
\end{enumerate}

In summarizing the evaluations conducted using PPS, AER, and AOP, we observe that the four curvatures – LRC, FRC, BFC, and ORC – exhibit comparable performance in community detection. None of the curvatures consistently outperforms the others across all metrics and all pairs of $p_1,p_2$, indicating a balanced landscape of effectiveness. 

However, when considering computational efficiency, LRC and FRC emerge as the fastest, both offering $\OO(mn)$ complexity. The crucial difference is that FRC is not scale-free, but LRC boasts this advantageous property, making it particularly suitable for analyzing large-scale networks where scalability is key. This distinction positions LRC as an ideal candidate for our proposed preprocessing method. Nevertheless, it is important to note that if specific scenarios or requirements strongly favor other curvatures, our preprocessing approach remains adaptable and can be effectively applied in a broader context.

Moving forward, the next section will focus on the application of LRC, leveraging its efficiency and scale-free nature, to real datasets. This will provide empirical insights into its practical utility and effectiveness in real-world network analysis scenarios.

\section{Application}

In this section, we evaluated the impact of our proposed LRC preprocessing method on the performance of various community detection algorithms using four real-world datasets with known community structures. We begin our analysis with a smaller network, the NCAA Football League Network, to demonstrate the impact of our preprocessing method in a more controlled setting. To this end, we compared the Adjusted Rand Index (ARI) and Adjusted Mutual Information (AMI) scores before and after applying our preprocessing technique, utilizing four representative community detection models: Label Propagation, Leiden, Girvan-Newman, and Walktrap. These models were chosen for their effectiveness and widespread use in community detection, as noted in the existing literature \citep{fortunato2016community}.

\subsection{NCAA Football League network}
The NCAA football Division I games schedule for the 2000 season \citep{girvan2002community} is represented in these graph data. It consists of 115 nodes, representing college football teams, and 613 edges, corresponding to the regular season games played between these teams. The dataset identifies 12 ground-truth communities or conferences. Since teams tend to play more frequently within their own conference, this network clearly exhibits a community structure. \Cref{tab:NCAA} below illustrates the performance improvement of various community detection algorithms through the application of our preprocessing method.

\begin{table}[h!]
\centering
\begin{tabular}{||c| c c c c||} 
 \hline
 &\multicolumn{4}{|c||}{Algorithms} \\
 \hline
 Scores & Label Propagation & Leiden & Girvan-Newman & Walktrap \\ [0.5ex] 
 
 \hline\hline
 ARI: before & 0.75 & 0.81 & 0.84 & 0.82 \\ 
 \hline
 ARI: after & {\bf 0.89} & {\bf 0.89} & {\bf 0.87} & {\bf 0.89} \\
 \hline
 AMI: before & 0.85 & 0.88 & 0.91 & 0.86 \\
 \hline
 AMI: after & {\bf 0.93} & {\bf0.93} & {\bf0.92} & {\bf0.93} \\
 \hline
 \hline
 Complexity & $\OO(m)$ & $\OO(nm)$ & $\OO(nm^2)$ & $\OO(n)$\\
 \hline 
\end{tabular}
\caption{Community detection algorithms evaluation for NCAA Football League network}
\label{tab:NCAA}
\end{table}

The results clearly show an improvement in both ARI and AMI scores after applying our preprocessing method across all four community detection algorithms. Notably, the Label Propagation algorithm, which initially had the lowest ARI (0.75) and AMI (0.85), significantly improved to 0.89 and 0.93, respectively, after preprocessing. This enhancement elevates it to one of the top-performing algorithms in this context. In fact, after preprocessing, all algorithms exhibit very similar scores, suggesting that our method simplifies the community detection problem by making the community structures more distinct and evident. This homogenization of performance implies that, with effective preprocessing, the choice of community detection algorithm becomes less critical, as the clarified network structure facilitates more accurate community detection across different methods.

Importantly, the inclusion of computational complexity in our analysis (as shown in the last row of the table) provides further insight into the selection of an optimal algorithm. Algorithms such as Label Propagation and Walktrap, with their lower computational complexities of $\OO(m)$ and $\OO(n)$ respectively, become attractive options. This highlights another significant advantage of our preprocessing method – it not only improves the accuracy of community detection, but also enhances overall efficiency by enabling the use of faster algorithms without compromising on performance.

After evaluating the NCAA Football League Network, we extend our analysis to three larger-scale networks. These networks pose additional challenges, particularly in terms of computational efficiency. Moreover, they often exhibit mixed membership, where nodes can belong to multiple communities, diverging from the unique community structures seen in smaller and simpler networks like the NCAA dataset. 

Given these differences, we shift our focus to algorithms better suited for these conditions. For larger networks, we use Angel, Ego, K-clique, and Speaker-Listener Label Propagation Algorithm (SLPA). These replacements are due to the suitability of these algorithms in handling large-scale networks and their capability to address mixed membership scenarios.

Furthermore, the ARI and AMI scores are less effective for evaluating community detection in networks with mixed memberships. Therefore, we utilize the F1 score, a well-established metric in such scenarios, which combines the precision and recall of the detected communities to provide a balanced measure of a method's accuracy and is particularly useful in networks where a node can be part of multiple communities~\citep{hollocou2018multiple}.

\subsection{DBLP collaboration network}

The DBLP computer science bibliography co-authorship network \citep{yang2012defining} is another dataset we explored. Each node represents a researcher, and each edge signifies a collaborative paper. The network comprises 317,080 nodes and 950,059 edges. The results, as shown in the table below, demonstrate that the LRC preprocessing method aids in detecting community structures in more complex networks.

\begin{table}[h!]
\centering
\begin{tabular}{||c| c c c c||} 
 \hline
 &\multicolumn{4}{|c||}{Algorithms} \\
 \hline
 Scores & Angel  & Ego & K-clique & SLPA \\ [0.5ex] 
 \hline\hline
 F1: before & 0.284   & 0.317 & 0.276 & 0.211\\ 
 \hline
 F1: after & {\bf0.452}   & {0.386} & {0.420} & {0.371}\\
 \hline
   \hline
Time: before  & 260.17  & 831.87 & 40.50 & 1024.12\\
 \hline 
 Time: after& {\bf 180.39}  & 184.88 & 111.60 & 2349.36 \\
 \hline 
\end{tabular}
\caption{Community detection algorithms evaluation for DBLP network}
\label{table:DBLP}
\end{table}

Before delving into the results presented in Table \ref{table:DBLP}, it is important to note that due to the complex nature of the community detection process in large-scale networks, the traditional Big O notation for computational complexity is not reported here. Instead, we focus on the actual runtime of each algorithm, providing a more practical measure of efficiency in real-world applications. The `Time: before' represents the runtime (in seconds) of each community detection algorithm when applied directly to the raw network data, while `Time: after' encompasses the total time, which includes calculating LRC, identifying the threshold, removing edges based on this threshold, and rerunning the same community detection algorithm on the processed network. This approach ensures a fair comparison, as it accounts for all steps involved in our preprocessing method.

Table 3 reveals a notable improvement in the F1 scores for each algorithm after integrating our preprocessing method to DBLP collaboration network. Furthermore, this improvement in performance does not come at the cost of reduced efficiency; in fact, the Angel algorithm demonstrates increased processing speed post-preprocessing, even with these additional preprocessing steps, highlighting the efficiency of our method in complex network environments.

\subsection{Amazon product co-purchasing network}

This dataset represents the co-purchasing patterns of products on Amazon \citep{yang2012defining}. The nodes symbolize products, and the edges indicate co-purchases by Amazon customers. The network includes 334,863 nodes and 925,872 edges. As with the previous datasets, the application of the LRC preprocessing method significantly enhanced the results of various community detection algorithms, as illustrated in the table below.

\begin{table}[h!]
\centering
\begin{tabular}{||c|  c c c c||} 
 \hline
 &\multicolumn{4}{|c||}{Algorithms} \\
 \hline
 Scores & Angel & Ego & K-clique & SLPA \\ [0.5ex] 
 \hline\hline
 F1: before & 0.368 & 0.371 & 0.387 & 0.345 \\ 
 \hline
 F1: after & {0.463} & {0.444} & {\bf0.482} & {0.483} \\
 \hline
   \hline
Time: before  & 159.52 & 1000.85 & 42.40 & 2380.61\\
 \hline 
 Time: after  & 139.61& 629.56 & {\bf 85.73} & 3911.19\\
 \hline 
\end{tabular}
\caption{Community detection algorithms evaluation for the Amazon network}
\label{table:Amazon}
\end{table}

In line with the results of DBLP collaboration network, our method improved the performance of each community detection algorithm. Notably, this consistency underscores the robustness of our preprocessing approach across different types of large-scale networks.

\subsection{YouTube social network}

This dataset represents the social network on YouTube \citep{mislove-2007-socialnetworks}. Nodes symbolize users, and edges indicate friendship such as subscription between YouTube users. The network includes 1,134,890 nodes and 2,987,624 edges. The results show that the implementation of the LRC preprocessing method greatly improved the results of multiple community detection algorithms in line with prior datasets.

\begin{table}[h!]
\centering
\begin{tabular}{||c| c c c c||} 
 \hline
  &\multicolumn{4}{|c||}{Algorithms} \\
 \hline
 Scores & Angel & Ego & K-clique & SLPA \\ [0.5ex] 
 \hline\hline
 F1: before & 0.063 & 0.22 & 0.066 & 0.093\\ 
 \hline
 F1: after & 0.282  & {\bf0.44} & 0.216 & 0.429\\
 \hline
   \hline
 Time: before  & 972.74  & 143.09 & 9029.98 & 117.23\\
 \hline 
 Time: after  & 67.840  & {\bf 129.63}& 131.75 & 218.29\\
 \hline 
\end{tabular}
\caption{Community detection algorithms evaluation for the YouTube network}
\label{table:Youtube}
\end{table}

Table \ref{table:Amazon} showcases the effectiveness of our preprocessing method in the YouTube social network. While the performance boost is apparent across all algorithms, the Ego and SLPA algorithms stand out for their marked improvements in F1 scores. This result diverges slightly from other large networks, as K-clique is not the fastest method here. Nevertheless, our method consistently enhances the overall performance of community detection, particularly benefiting faster analysis methods.

In conclusion, across all three large networks analyzed – DBLP, Amazon, and YouTube – our LRC preprocessing method consistently enables at least one community detection algorithm to achieve the best or near-best performance scores, while maintaining impressive efficiency with runtimes under 200 seconds. For instance, the Angel algorithm for the DBLP network, K-clique for the Amazon network, and Ego for the YouTube network each emerged as top performers in their respective datasets. This is particularly noteworthy given the substantial size of these networks. Such results underscore the exceptional effectiveness and efficiency of our preprocessing approach in handling complex, large-scale network data, making it a highly valuable tool in the field of network analysis.

\section{Discussion and future work}
In this work, we have focused on network curvature and its applications in community detection. Our key contribution is the proposal of the Lower Ricci Curvature (LRC), a scalable and scale-free discrete curvature designed specifically for network analysis. Alongside this, we have developed an LRC-based preprocessing method that has shown potential in enhancing the performance of established community detection methods. This assertion is backed by both simulations and real-world applications, including analyses of large-scale networks such as Amazon, DBLP, and YouTube. Moreover, the LRC framework is adaptable and can be straightforwardly extended to weighted networks. Looking forward, several promising directions for extending this research are evident.

\begin{enumerate}
\item Extension to directed graphs: Extending LRC to directed graphs opens up numerous possibilities for analysis in various fields. Directed graphs are crucial in representing asymmetric relationships, such as citation networks in academia, where the directionality of citations plays a significant role~\citep{newman2001structure}, or in web link structures where the direction of links implies a flow of information~\citep{kleinberg1999web}. Adapting LRC to account for the directionality in such networks can provide more nuanced insights into their structural and community dynamics. 

\item Application to hypergraphs: Hypergraphs, which involve higher-order interactions beyond pairwise connections, present an exciting frontier. For instance, in collaborative environments like multi-author scientific publications~\citep{taramasco2010academic} or gene co-expression ~\citep{tran2012hypergraph}, interactions are inherently multi-dimensional. Extending LRC to hypergraphs could yield a deeper understanding of these complex relational structures and the underlying community formations.

\item Deeper theoretical investigation of LRC: There is ample scope for exploring the theoretical aspects of LRC. Investigating the asymptotic behavior of the mixing components in different network models, such as the SBM, could provide valuable theoretical insights. Additionally, a thorough analysis of the three scores (PPS, APW, and AOP) under various network models could deepen our understanding of LRC's effectiveness and limitations in community detection.
\end{enumerate}



\bibliographystyle{chicago}
\bibliography{ref.bib}

\begin{appendix}
\section{Proof of \Cref{prop:bounds} and \Cref{cly:Cheeger}}
$\LRC\leq \BFC$ directly follows from definition, as $\BFC-\LRC=\Delta\geq$. The inequality $\BFC\leq \ORC$ is derived from Theorem 2 in \cite{topping2021understanding}. \Cref{cly:Cheeger} is a direct consequence of \Cref{prop:bounds} combined with Corollary 3 and Proposition 5 from \cite{topping2021understanding}.

\section{Additional experimental details}
\subsection{Hyperparameters for community detection}
All algorithms implemented in this paper are from Python package CDlib \citep{rossetti2019cdlib}. The hyperparameters are as follows:

\begin{enumerate}
    \item NCAA Football League network
    
{\bf Label Propagation:} NA.

{\bf Leiden:} Initial membership = None, weights= None.

{\bf Girvan-Newman:} Level=10.

{\bf Walktrap:} NA.

    \item DBLP collaboration network
    
{\bf Angel:} Threshold = 0.5, minimum community size = 3.

{\bf Ego-networks:} Level = 1.

{\bf K-clique:} $K = 3$.

{\bf SLPA:} $t=20, r=0.1$.

    \item Amazon product co-purchasing network

{\bf Angel:} Threshold = 0.5, minimum community size = 3.

{\bf Ego-networks:} Level 1.

{\bf K-clique:} $K= 3$.

{\bf SLPA:} $t=20, r= 0.1$.

    \item YouTube social network

{\bf Angel} Threshold = 0.5, minimum community size = 3.

{\bf Ego-networks} Level 1. 

{\bf K-clique} $K=7$.

{\bf SLPA} $t =20, r =0.1$.
    
\end{enumerate}

\subsection{Code and data availability}
All codes can be found in \url{https://github.com/parkyunjin/LowerRicciCurv.git}

\noindent The four real datasets used in this paper can be downloaded in the following websites:

\begin{enumerate}
\item NCAA Football League network: \url{https://websites.umich.edu/~mejn/netdata/} under ``American College football".
\item DBLP collaboration network: \url{https://snap.stanford.edu/data/com-DBLP.html}
\item Amazon product co-purchasing network: \url{https://snap.stanford.edu/data/com-Amazon.html}
\item YouTube social network: \url{https://snap.stanford.edu/data/com-Youtube.html}
\end{enumerate}

\end{appendix}

\end{document}